\documentclass[preprint]{aastex}
\slugcomment{Submitted to ApJL}
\shortauthors{Zheng et al.}
\begin{document}

\title{The initiation of a solar streamer blowout coronal mass ejection arising from the streamer flank}
\author{Ruisheng Zheng, Yao Chen, and Bing Wang}
\affil{Shandong Key Laboratory of Optical Astronomy and Solar-Terrestrial Environment, School of Space Science and Physics, Institute of Space Sciences, Shandong University, Weihai, Shandong, 264209, People's Republic of China; ruishengzheng@sdu.edu.cn; yaochen@sdu.edu.cn}

\begin{abstract}
Streamer blowout (SBO) coronal mass ejections (CMEs) represent a particular class of CMEs that are characterized by a gradual swelling of the overlying streamer and a slow CME containing a flux-rope structure. SBO CMEs arising from the streamer flank fall into a special category of SBO CMEs involving three lower arches under the higher streamer arcade. However, the initiation mechanism for this special category of SBO CMEs remains elusive, due to the observational limitations. Here we report critical observations of a SBO CME associated with the eruption of a polar crown filament that originated from the streamer flank. The filament slowly rose toward the solar equator with the writhing motion, and underwent a sudden acceleration before its eruption. Interestingly, during the rising, the filament fields experienced gradual external reconnections, which is evidenced by the dip-shaped bottom of the enveloping flux-rope structure changing from a smooth concave, the slow inflows ($\sim$1.8 km s$^{-1}$) from both the filament fields and the coronal loops beneath, and the persistent brightenings around the interface between the filament fields and the coronal loops beneath. The newly formed lower loops at the filament source and the Y-shaped structure in the stretched tail fields indicate the internal reconnections for the filament eruption. The clear signatures of the external and internal reconnections shed light on the initiation mechanisms of SBO CMEs.

\end{abstract}


\section{Introduction}
Coronal mass ejections (CMEs) appear as bright transient features propagating in white-light coronagraph observations, and are generally believed to be the most violent eruptions in the solar system. CMEs release large amounts of magnetized plasma and energy from closed magnetic fields in the inner corona, which can drive space weather near the Earth. Hence, the origin and the initiation of CMEs are key to predicting potential space weather catastrophes. Currently, popular CME-triggering mechanisms include the breakout model (Antiochos et al. 1999) in which the sheared core fields inflate and reconnect with the overlying arcades (see Chen et al. (2016) for latest observations on breakout reconnection), the tether-cutting model (Moore et al. 2001) in which the reconnections between sheared arcades build and release a flux rope, and MHD flux-rope instabilities such as the torus/kink instability models (T{\"o}r{\"o}k et al. 2004; Kliem \& T{\"o}r{\"o}k, 2006; Chen et al. 2007; Kliem et al. 2010). The details of the initiation mechanisms for CMEs can be found in some comprehensive reviews (Forbes et al. 2006; Chen 2011; Shibata \& Magara 2011; Aulanier 2014; Janvier et al. 2015).

During the propagation from the inner corona to interplanetary space, CMEs usually encounter and interact with streamers that are tracers of the global coronal magnetic field configuration and the interfaces between opposite-polarity open fluxes (Low 1994; Chen et al. 2010). As a particular class of CMEs, streamer blowout (SBO) CMEs were first named by Sheeley et al. (1982) and characterized by a gradual swelling of the overlying streamer and an embedding flux-rope structure. SBO CMEs have a slow average speed of $\sim$390 km s$^{-1}$, and are long-lived events with an average duration of 40.5 hr, according to the statistics results of 909 events in 1996--2015 (Vourlidas \& Webb 2018). SBO CMEs do not correlate with sunspot number and dominate during solar minimum. The characteristics suggest that SBO CMEs arise from extended polarity inversion lines (e.g., quiet-Sun and polar crown filaments) away from active regions (ARs), and are triggered by flux-rope instabilities or magnetic reconnections that release magnetic energy accumulated by differential rotation (Vourlidas \& Webb 2018).

Successful eruptions arising from the streamer flank compose a special subset of CMEs in which the erupting core field containing a flux-rope structure gives rise to an SBO CME. If the eruption from the streamer flank is on a small scale, the CME will appear as a narrow streamer puff that was named by Bemporad et al. (2005). It has been suggested that the SBO CME arising from the streamer flank requires a special magnetic configuration consisting of three independent lower arches and the overlying large streamer arcade, and its initiation always involves the external and internal magnetic reconnections of the erupting fields (Bemporad et al. 2005; Moore \& Sterling 2007). On the other hand, due to the initial location under one streamer leg, the strong influence of the streamer open fields can make a successful eruption experience a nonradial movement before traveling out along the radial streamer (Jiang et al. 2009; Yang et al. 2012). This nonradial trajectory of the eruptions of high-latitude filaments naturally arises from the multipolar flux systems, and the deflection toward the streamer belt (heliospheric current sheet) has been confirmed by the numerical simulations (T{\"o}r{\"o}k et al. 2011; Zuccarello et al. 2012; Lynch \& Edmondson 2013.

However, the evolution of SBO CMEs in the lower corona is rarely reported, and the initiation mechanism of SBO CMEs still remains unresolved, which is likely due to observational limitations. Here we report a SBO CME on 2019 August 12-13 that arose from the streamer flank with clear observational signatures of magnetic reconnections, and sheds light on the initiation mechanism for SBO CMEs.

\section{Observations}
The SBO CME was well captured in the east limb by the Large Angle and Spectrometric Coronagraph (LASCO; Brueckner et al. 1995) C2 and C3. The filament eruption in the lower corona, as the precursor to the CME, was clearly recorded by observations from the Atmospheric Imaging Assembly (AIA; Lemen et al. 2012) on board the Solar Dynamics Observatory (SDO; Pesnell et al. 2012). We mainly employ the AIA passbands of 193~{\AA} (Fe XII, $\sim$1.6 MK), 171~{\AA} (Fe IX, $\sim$0.6 MK), and 304~{\AA} (He II, $\sim$0.05 MK). Each AIA image covers the solar full disk up to 0.5 $R_\odot$ above the limb, with a pixel resolution of 0.$"$6 and a cadence of 12 s. In addition, we also check the filament eruption in the Extreme Ultraviolet Imager (EUVI; Howard et al. 2008) on board the spacecraft of Solar-Terrestrial Relations Observatory (STEREO: Kaiser et al. 2008) ahead (-A) of the Earth.

We also capture the faint trajectories of the rising filament and the enveloping coronal structures using the persistence mapping technique (Thompson \& Young 2016) in which the value at each pixel is set as the maximum at this same pixel among a set of images. The erupting structures are also identified by the processed normalized-intensity images, in which the pixel intensity at the center time of every two hours is normalized by values at the same pixel in the 2 hr sequential images. The magnetic field lines are extrapolated with the potential field source surface (PFSS; Schrijver {\&} De Rosa 2003) package SolarSoftWare. The kinematics of the filament and enveloping coronal structures are obtained by the time-slice approach.

\section{Results}
\subsection{SBO CME kinematics and source-region configuration}
The evolution of the SBO CME is shown in LASCO C2 and C3 in Figure 1 and its animation. Before the eruption, a streamer (the blue arrow) existed near the equator in the east limb (panel (a)). Following the eruption, the CME arose beneath the streamer, and first appeared in LASCO C2 at $\sim$21:12 UT on August 12. The CME exhibited a typical three-part structure including an obvious expanding flux-rope structure (the green arrows) that propagated radially along the streamer, and the gradual swelling of the streamer (panels (b)-(c)). The particular signatures in coronagraph observations clearly demonstrate the occurrence of a classical SBO CME. In the plots of heights and velocities from the measurements of the LASCO CME catalog{\footnote{\url{https://cdaw.gsfc.nasa.gov/CME\_list/UNIVERSAL/2019\_08/htpng/20190812.211209.p080g.htp.html}}} (panels (d)-(e)), the CME had a central position angle of 80$^\circ$, an angular width of $\sim$37$^\circ$, and a linear speed of $\sim$224 km s$^{-1}$. Based on a second-degree fit, the CME had an acceleration of $\sim$7 m s$^{-2}$, and achieved a final speed of $\sim$468 km s$^{-1}$ at $\sim$16:12 UT on August 13 at the edge of the LASCO C3 field, after a duration of $\sim$19 hr for the streamer swelling.

The coronal fields associated with the CME are checked by the PFSS extrapolation method, and the extrapolated field lines are shown from the perspectives of SDO and STEREO-A in Figure 2. The open field lines (green and red lines) and large arcades (higher blue lines) confirm the existence of the streamer. Beneath the large arcades of the streamer, there are three sets of small lower arches (yellow lines). Note that the central arches were narrow (red arrows), and the north arches covered a polar crown filament that was the precursor to the SBO CME. At the extrapolated time ($\sim$06:04 UT on August 12), the filament rose higher in the limb viewpoint (the white arrow in panel (a)), and it only left faint projection traces in the disk viewpoint (the white arrow in panel (b)).

\subsection{Nonradial trajectory of the filament eruption}
The filament underwent a long-term gradual ascent that started from the solar rear before August 11. Figure 3 and its animation show the rise of the filament and the associated coronal structures in AIA images. In AIA 304~{\AA}, the rising filament (white arrows) first exhibited a long north-south arch (panel (a)), and then transformed into a writhed inverse-$\gamma$-shaped structure with two crossing legs (panel (b)), and finally move southward. It indicates a nonradial movement of the rising filament, and its south leg was clearly stretched southward (panel (c)). In AIA 171~{\AA} (panels (e)-(g)), the ambient coronal loops outlined clearly the bottom of a flux rope containing the filament (red arrows), and the flux rope also slowly drifted with the movement of the filament. In persistence images in AIA 304 and 171~{\AA} from 00:00 to 24:00 UT on August 12 (panels (d) and (h)), it is very clear for the movement trajectories (white and red arrows) of the rising filament and the surrounding coronal structures. Along a chosen direction (S1; dashed lines in panels (d) and (h)), the time-distance intensity plots in AIA 304 and 171~{\AA} (panels (i)-(j)) show that the rising speed of the filament and enveloping coronal structures was as low as $\sim$3 km s$^{-1}$ (the blue dotted line) before 12:00 UT on August 12 (the dashed vertical line), and accelerated to $\sim$8 km s$^{-1}$ (green dotted lines) before they left the AIA field.

%In the perspective of STEREO-A, the faint filament was identified in the running difference images (panels (a)-(c) of Figure 4). The filament first moved southeastward, and became fainter and fainter (panel (a)). Interestingly, nearly 10 hours later, the filament appeared clear again, but drifted southwestward (panels (b)-(c)). This indicates an apparent deflection of filament movement.

Interestingly, the nonradial rising filament interacted with the coronal fields beneath, and finally erupted, which is shown in Figure 4 and its animation. Though the interaction was gradual and subtle, the smooth bottom of the flux-rope structure obviously became dip-shaped (red arrows), and some lower loops (blue arrows) that connected the filament source and the sites under the dip-shaped side in AIA 171~{\AA} (panels (a)-(b)) formed. In the time-distance intensity plot in AIA 171~{\AA} (panel (c)) along the dotted line (S2) in panel (d), it is clear that the magnetic fields of drifting flux rope and the coronal loops beneath got close to each other from $\sim$12:00 UT on August 12 (the vertical line) for more than 10 hr. The flux-rope fields sank with a speed of $\sim$1.8 km s$^{-1}$ (the blue dotted line), and the coronal loops beneath ascended with a speed of $\sim$1.7 km s$^{-1}$ (the red dotted line). The approach of the fields of the drifting flux rope and the coronal loops beneath is likely an indicator of the inflows of magnetic reconnections. In the base-ratio-difference image in AIA 171~{\AA} (panel (d)), the fields of the flux rope and the coronal loops beneath became very bright around the dip-shaped bottom (red arrows), which was possibly due to the heating from the magnetic reconnection.

Following the nonradial movement and the sinking of the enveloping fields, there appeared a faint Y-shaped structure (the blue arrow) in the tail of the stretched filament fields, which is better seen in the normalized-intensity image in AIA~{\AA} at $\sim$00:21 UT on August 13 (panel (e)).  On the other hand, some lower arches (green arrows) newly formed at the filament source in the base-ratio-difference images in AIA 171 and 193~{\AA} (panels (f)-(g)). The Y-shaped structure and the newly formed lower arches are likely productions of the filament eruption, and are shown clearly in the insets (blue dotted boxes) in panels (e)-(g). In addition, the brightenings of the filament fields lasted for some hours (red arrows in panels (f)-(g)), which possibly implies a succession of magnetic reconnections.

\section{Conclusions and Discussion}
The white-light observations from LASCO captured a bright CME (Figure 1), and the PFSS extrapolated field lines confirm the configuration of a streamer enwrapping three sets of arches beneath, and a polar crown filament lies under the north arches in the streamer flank (Figure 2). As the precursor to the CME, the filament eruption in the lower corona is well recorded in AIA images. The filament rose slowly toward the solar equator with a writhing motion, and the ascent speed of the enveloping fields accelerated from $\sim$3 to $\sim$8 km s$^{-1}$ (Figure 3). The obvious signatures, a gradual swelling of the overlying streamer for $\sim$19 hr and a well-structured flux rope embedding in the CME, are consistent with the characteristics of typical SBO CMEs (Vourlidas \& Webb 2018).

Interestingly, the SBO CME originated from a polar crown filament that located in the streamer flank far from ARs. In the special configuration, the overlying streamer is strong enough to make the erupting core rise nonradially (Jiang et al. 2009; Yang et al. 2012), and the nonradial movement can lead the filament fields to encounter and interact with neighboring coronal fields. The initiation mechanism for an eruption arising from the streamer flank was suggested to involve external and internal reconnections 15 yr ago (Figure 3 in Bemporad et al. 2005), but the observational evidence of external reconnections is rarely reported, to our knowledge. In a few cases of the similar eruptions arising from the streamer flanks (Moore \& Sterling 2007; Yang et al. 2012) only coronal dimmings were found as the indirect signature of external reconnections. In this Letter, the external reconnections between the rising filament fields and coronal loops beneath are supported by the deformation of the bottom shape of the enveloping flux rope from the smooth concave to the dip (the red arrow in Figure 4(a)-(b)), the very slow inflows ($\sim$1.8 km s$^{-1}$) that make the filament fields and coronal loops beneath get close to each other (Figure 4(c)), and the persistent brightenings around the interface between the filament fields and coronal loops beneath (red arrows in bottom panels of Figure 4). The signatures of associated coronal dimmings were too weak to identify with both SDO and STEREO-A's perspectives, which is possibly because of the weakness of the eruption. Following external reconnections, the filament finally erupted, and there appeared a Y-shaped structure implying a detachment of the stretched erupting core and the post-eruption loops in the filament source (blue and green arrows in bottom panels of Figure 4) as a result of internal reconnections of the stretched filament fields.

Both external and internal reconnections occurred. What did the external reconnection contribute to the successful release of the SBO CME? Jiang et al. (2009) reported that the eruption of an AR filament from the streamer flank directly blew the large streamer arcades open, and formed a narrow CME without external reconnections. It likely indicates that enough energy from the erupting core fields is the key to break through the large closed arcade of the overlying streamer, and AR filaments always store enough magnetic energy. The event duration is one important parameter for SBO CMEs, and follows a trend toward long durations for less energetic events. For this case, the short duration of 19 hr indicates an energetic event. Moreover, the polar crown filament exhibited a writhed inverse-$\gamma$-shaped structure during the ascent, which implies an amount of accumulated free energy and a kink instability before the rising. Kink instability can be crucial for triggering an eruption (T{\"o}r{\"o}k et al. 2004). However, the rising speed of filament fields remained slow ($\sim$3 km s$^{-1}$) during the writhing motion (panels (b) and (j) in Figure 3) and began to accelerate to a higher value ($\sim$8 km s$^{-1}$; dotted lines in Figure 3) nearly simultaneously with the beginning of the inflows (vertical dashed lines in Figures 3 and 4). This likely indicates the kink instability is not the key to trigger the eruption, and external reconnection plays an important role in the initiation of the SBO CME.

Therefore, based on extrapolated PFSS field lines (Figure 2) and AIA observations (Figure 3-4), we propose a possible scenario for the initiation of the SBO CME in the schematic (Figure 5). In the pre-eruption magnetic configuration (panel (a)), the streamer consists of open fields (green and red lines) and a large arcade (blue loops), and three lower arches (pink, black, and orange) underneath the large streamer arcade. The northernmost arches (pink) involve the rising filament fields rising from the steamer north flank. The open fields root in the north positive polar region and in the south negative polar region, respectively, checked by magnetograms from the SDO. From north to south, the polarities are positive, negative, positive, and negative (green pluses and red minuses). During the ascent (panel (b)), because of the guidance of the north open fields, the northernmost rising filament fields move southward and interact with the southernmost arches (orange). The interaction contributes to the external reconnections (the red cross) of the filament fields, and the external reconnections results in a newly formed central arch (the red loop) and a curved large arcade (the cyan loop). Due to the decrease of the confinement, the filament rises faster, the opposite field lines of the stretched filament tail quickly get closer, and internal reconnections (the green cross) occur. Consequently, the filament erupts, accompanying with the formation of a closed flux-rope structure (the pink circle) detached from the solar surface and a lower arch (the green loop) at the filament source (panel (c)). Moreover, after the complete detachment of the filament fields from the solar surface, the streamer arcades are stretched sequentially, and then another reconnection occurs between the approaching legs of streamer arcades, which transfer the flux into the CME from below. On the other hand, through the slow lift of the filament fields, the open fields are slowly blown out, and the outer layer of the large arcades becomes open into the solar wind, serving as an indicator of the flux loss from above (panels (b)-(c)).

The scenario of the configuration and evolution of the slow SBO CME are similar to that in Figure 3 of Bemporad et al. (2005). However, the pre-eruption magnetic topology (Figure 5(a)) of three polarity inversion lines (yellow dotted lines) and four closed flux systems (blue, pink, black, orange) is exactly the multipolar flux distribution of the magnetic breakout CME initiation scenario (Antiochos et al. 1999) that demands at least one coronal null point at the intersection of these distinct flux systems. The associated separatrix boundaries and the overlying coronal null point are indicated by the black dashed lines and the red circle, respectively, in Figure 5(a). This overlying null point will be distorted and stretched out into a current sheet when the pink arcade is energized with the presence of sheared or twisted filament channel fields. That current sheet will facilitate reconnection between the various flux systems. In fact, the external reconnection proposed by Bemporad et al. (2005) is essentially the breakout reconnection noted by Antiochos et al. (1999), and the internal or tether-cutting reconnection in Moore et al. (2001) has a similar topology to the flare reconnection underneath the erupting core in the standard CSHKP model (Carmichael 1964; Sturrock 1966; Hirayama 1974; Kopp \& Pneuman 1976).

All the results show that the initiation for the SBO CME rising from the streamer flank involves both external and internal reconnections. Are the external reconnections necessary for this SBO CME? There is an open question about how long this internal reconnection proceeds slowly (as a result of flux cancellation or converging flows) and builds up twist and energy in the flux-rope prior to the transition to the fast reconnection that causes the main acceleration and eruption. If there is no external reconnection, the polar crown filament will keep rising slowly, and accumulate much more magnetic energy via the differential rotation at a longer duration, and lead to internal reconnection at a higher altitude. As a result, the following CME likely leave weak or no low coronal signatures, making it a stealth CME (Robbrecht et al. 2009; Lynch et al. 2016) that is likely a subset of SBO CMEs. Hence, we suggest that external reconnections are important rather than necessary to induce the eruption onset in advance. Better observations of SBO CMEs are necessary to understand their initiation mechanisms.

\acknowledgments
SDO is a mission of NASA's Living With a Star Program. We gratefully acknowledge the use of data from SDO and from STEREO. This work is supported by grants NSFC 11790303 and U1731101.

\clearpage

\begin{figure}
\epsscale{0.95} \plotone{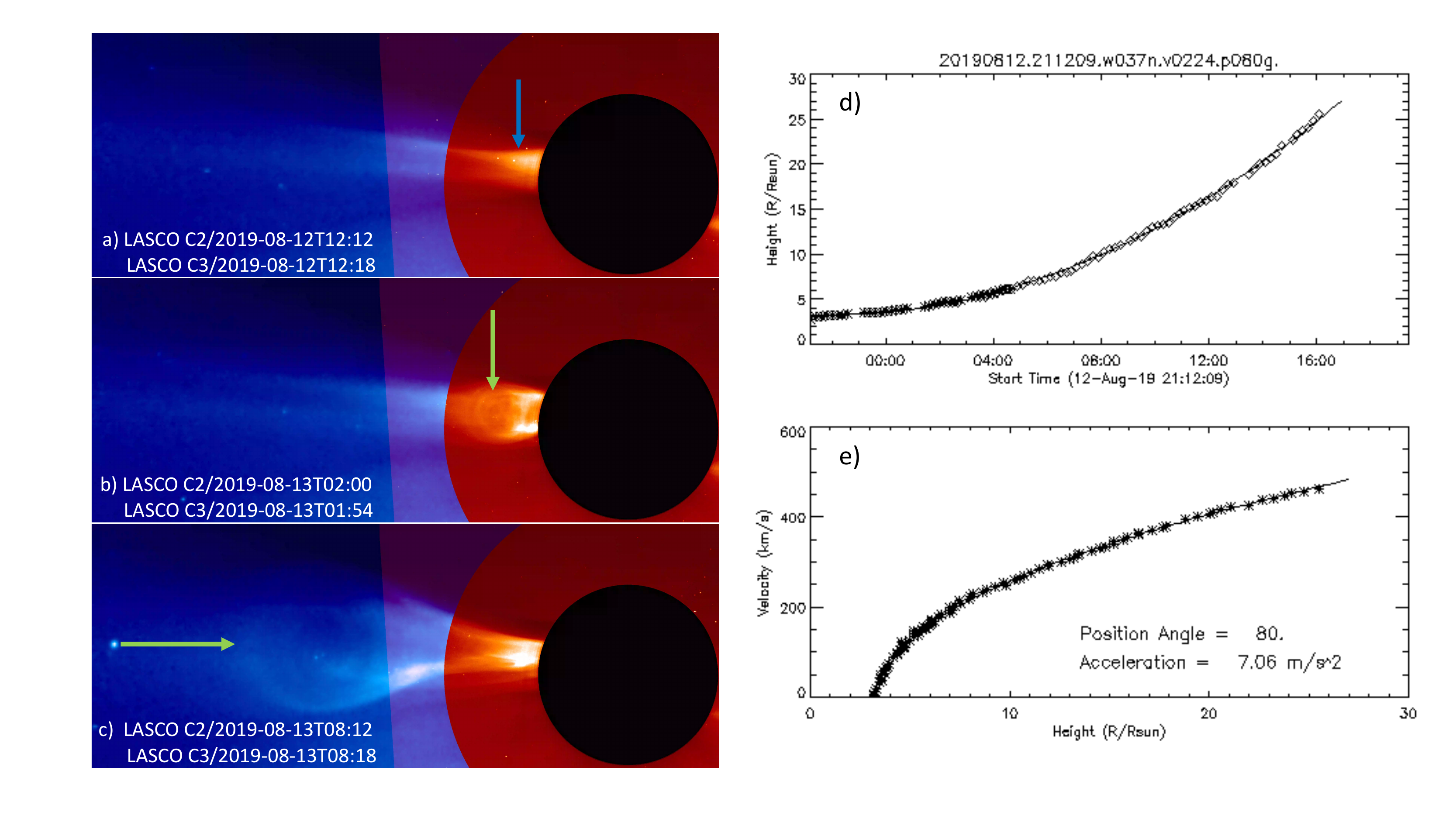}
\caption{SBO CME. (a)-(c) Combined images of LASCO C2 and C3 coronagraphs showing the streamer (the blue arrow) and the propagating flux-rope structure (green arrows) inside the SBO CME. (d)-(e) Height and velocity plots of the SBO CME from the LASCO CME catalog. The animation of the LASCO C2 and C3 images begins at $\sim$12:06 UT on August 12 and ends at $\sim$11:54 UT on August 13.
\label{f1}}
\end{figure}

\clearpage

\begin{figure}
\epsscale{0.9} \plotone{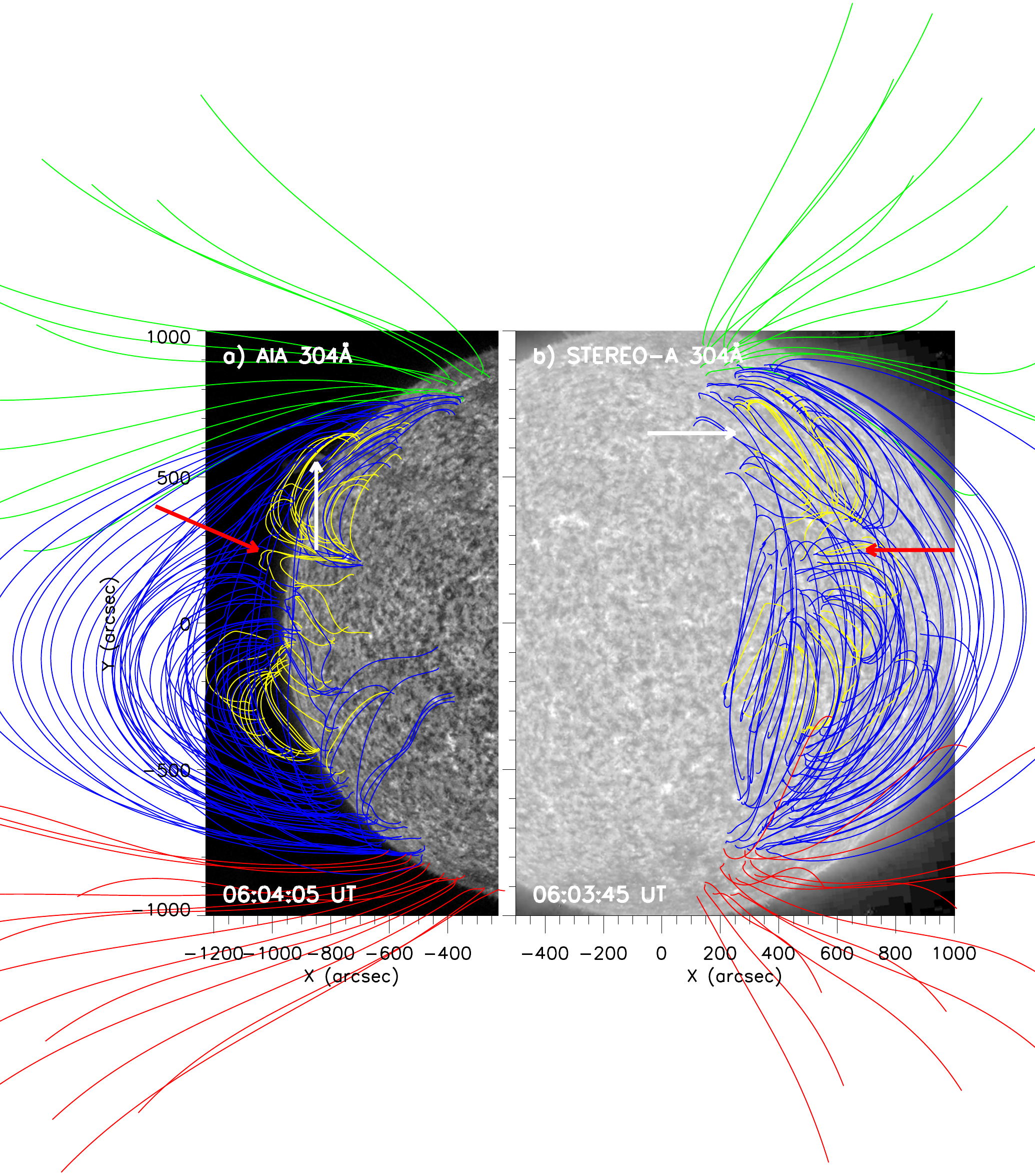}
\caption{Magnetic configuration of the streamer and the underlying higher arcades and lower arches. (a)-(b) The extrapolated PFSS field lines based on 304~{\AA} images from both SDO and STEREO-A. The white arrows show the rising filament, and the red arrows indicate the narrow arches adjacent to the filament arches.
\label{f2}}
\end{figure}

\clearpage

\begin{figure}
\epsscale{0.9} \plotone{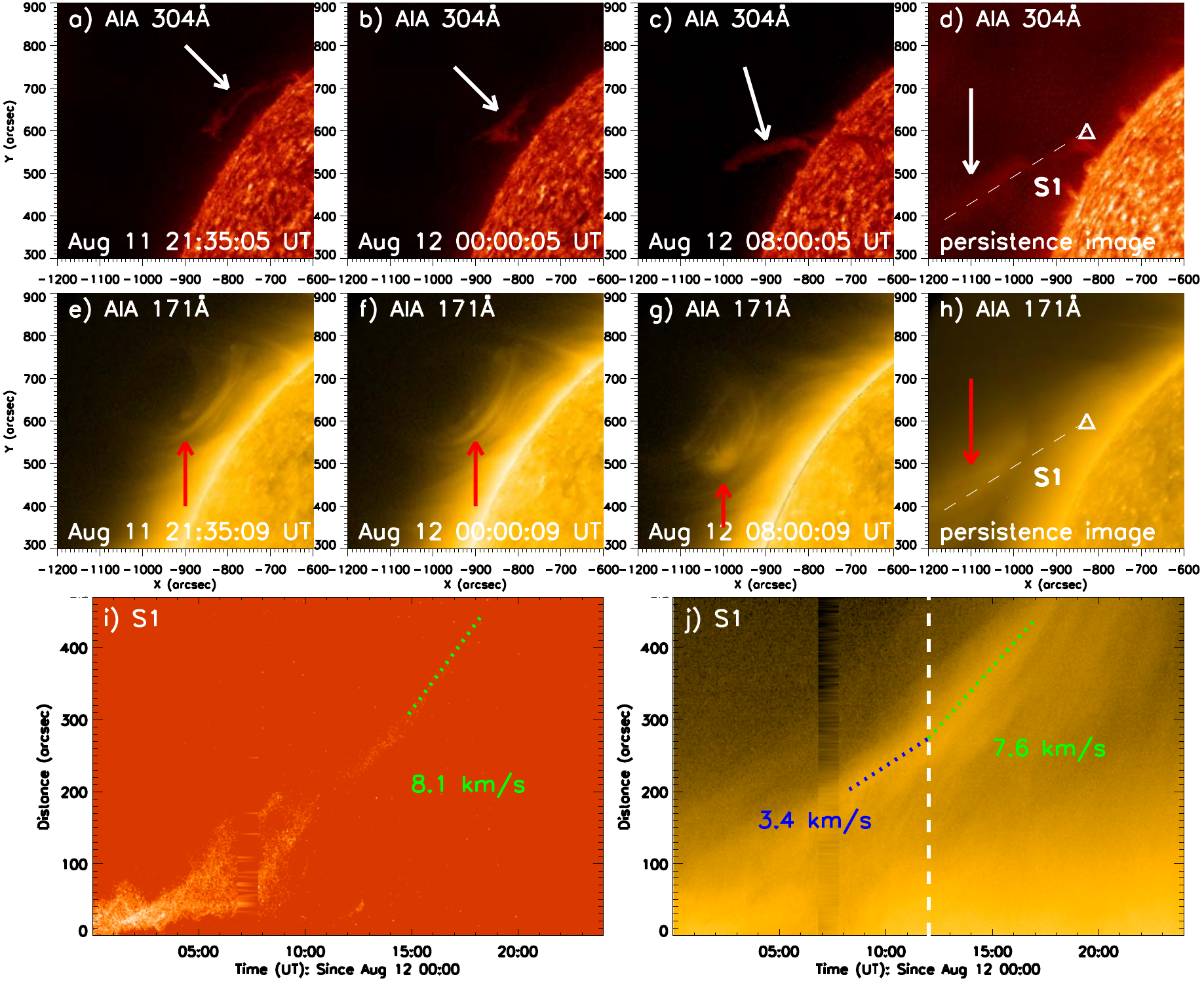}
\caption{Filament nonradial rise. (a)-(h) The nonradial movement of the filament (white arrows) and enveloping coronal loops (red arrows) in AIA 304 and 171~{\AA}. Their movement trajectories are shown in persistence images (panels (d) and (h)) between 00:00 and 24:00 UT on August 12. (i)-(j) Time-distance intensity plots along S1 (dashed lines with a triangle start in panels (d) and (h)) in AIA 304 and 171~{\AA}. The dotted lines are used to derive the attached speeds, and the vertical dashed line indicates 12:00 UT on August 12. The animation of AIA 304 and 171~{\AA} images begins at $\sim$21:05 UT on August 11 and ends at $\sim$23:06 UT on August 12.
\label{f3}}
\end{figure}

\clearpage

\begin{figure}
\epsscale{0.9}
\plotone{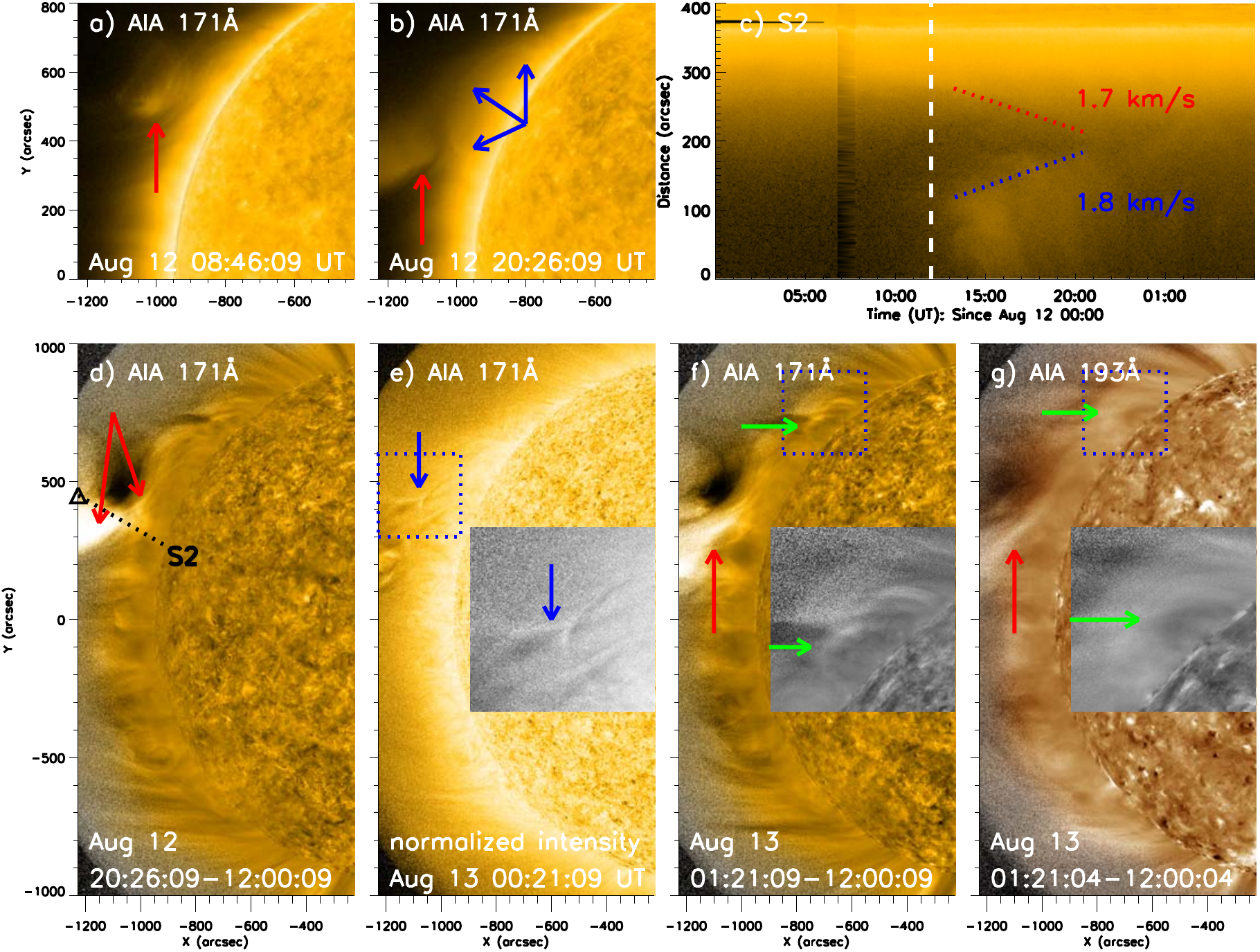}
\caption{Filament eruption. (a)-(b) AIA 171~{\AA} images showing the deformation (red arrows) of the coronal structures enveloping the filament and the newly formed lower loops (blue arrows) under the stretched filament fields. (c) Time-distance plot along S2 (the dotted line with a triangle start in panel (d)) in AIA 171~{\AA}. The dotted lines are used to derive the attached speeds, and the vertical dashed line indicates 12:00 UT on August 12. (d)-(g) Base-ratio-difference images in AIA 171 and 193~{\AA} and the normalized-intensity image in AIA 171~{\AA} showing the brightenings (red arrows), the Y-shaped structure (blue arrows), and post-eruption loops (green arrows). The fields of view of insets are indicated by blue dotted boxes. The animation of AIA 171 and 193~{\AA} images begins at $\sim$08:11 UT on August 12 and ends at $\sim$02:11 UT on August 13.
\label{f4}}
\end{figure}

\clearpage

\begin{figure}
\epsscale{0.7}
\plotone{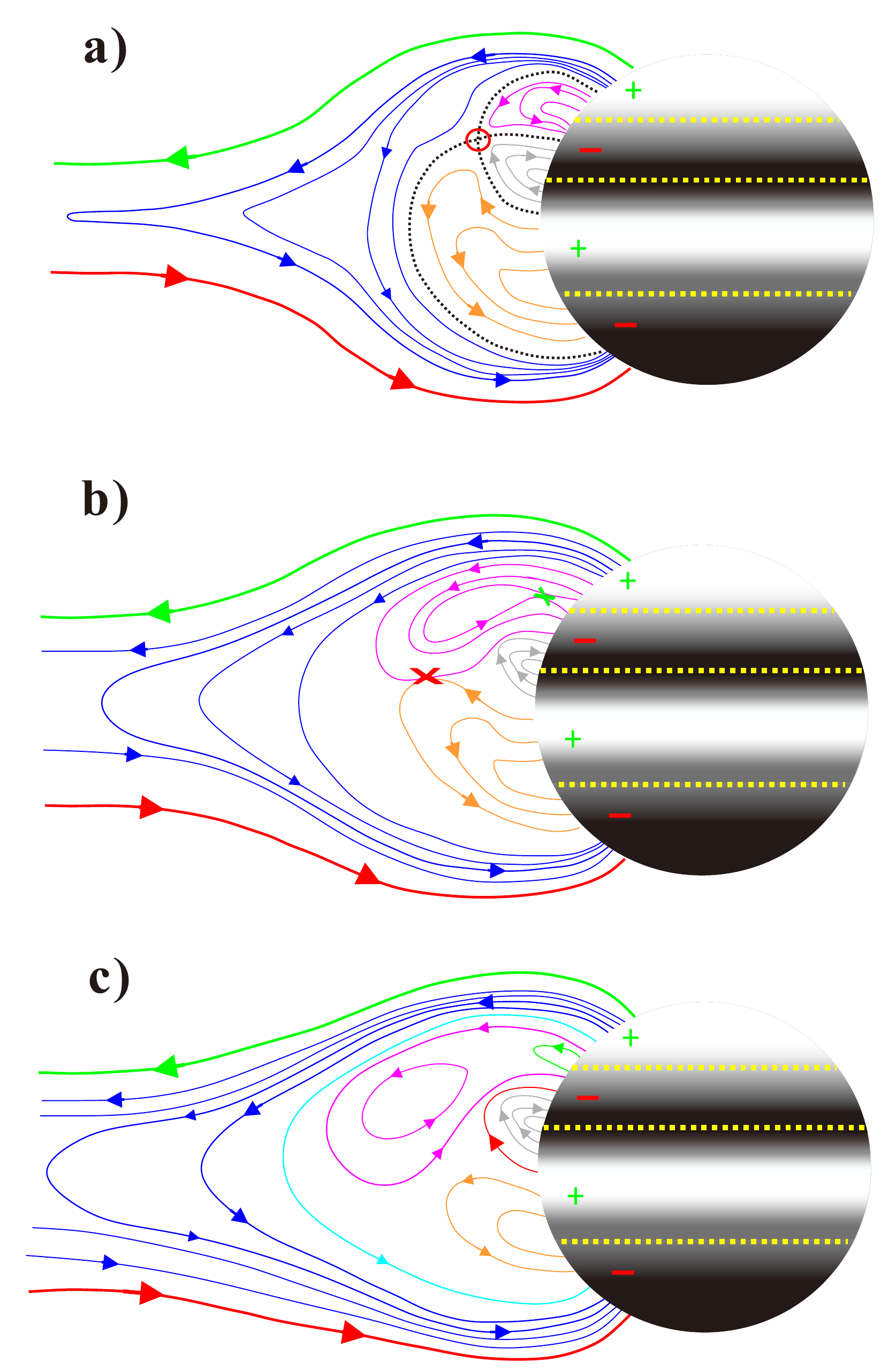}
\caption{Schematic of the initiation mechanism for the SBO CME, based on AIA 171~{\AA} images. (a) The pre-eruption magnetic configuration showing the open fields (green and red lines) and large arcade (blue) of the streamer, the rising filament fields (pink), the middle arches (black), and the southernmost arches (orange). (b) The external and internal reconnections (red and green crosses). (c) The magnetic configuration after the reconnection showing the newly formed post-eruption loop (the green loop), the erupting core (the pink closed loop), the middle lower arch (the red loop), and the large arcade (cyan). The green pluses and red minuses indicate the positive and negative polarities, and yellow dotted lines represent polarity inversion lines. The associated separatrix boundaries and the overlying coronal null point are indicated by the black dashed lines and the red circle, respectively.
\label{f5}}
\end{figure}

\end{document}